\documentclass[a4paper,twocolumn,preprintnumbers,citeautoscript,aps,floatfix]{revtex4-1}
\usepackage[utf8]{inputenc}
\usepackage{amsmath,amsfonts,amssymb,graphics}
\usepackage[normalem]{ulem}
\usepackage{graphicx}
\usepackage{units}
\usepackage{mathrsfs}
\usepackage{bbm}
\usepackage{pifont}
\usepackage{braket}
\usepackage{units}
\usepackage{footmisc}
\usepackage[dvipsnames]{xcolor}
\usepackage{mathtools}
\usepackage{xspace}
\usepackage{booktabs}
\usepackage{braket}

\usepackage[bookmarks=false,hyperfootnotes=false]{hyperref}

\definecolor{nicered}{rgb}{0.5,.0,.0}
\definecolor{darkblue}{rgb}{0,.1,.9}
\definecolor{lightblue}{rgb}{0,.1,.6}
 \definecolor{darkgreen}{rgb}{0.0,0.2,0.0}
\hypersetup{colorlinks, citecolor=darkgreen ,linkcolor=darkblue, urlcolor=lightblue}

\newcommand{\keV}{\ensuremath{\,\mathrm{keV}}\xspace}

\newcommand{\eV}{\ensuremath{\,\mathrm{eV}}\xspace}

\newcommand{\Eqref}[1]{Eq.~\eqref{#1}}
\newcommand{\Figref}[1]{Fig.~\ref{#1}}
\newcommand{\Tb}[1]{\ensuremath{\hat{T}_b^{#1}(k)}\xspace}
\newcommand{\TF}[1]{\ensuremath{\hat{T}_\mathrm{F}^{#1}(k)}\xspace}
\newcommand{\Tk}{\ensuremath{\mathcal{T}(k)}\xspace}
\newcommand{\Gb}{\ensuremath{\hat{\mathcal{G}}_b(r)}\xspace}
\newcommand{\kF}{\ensuremath{k_{\mathrm{F}}}\xspace}
\newcommand{\kU}{\ensuremath{k_{\mathrm{U}}}\xspace}

\newcommand{\Si}[1]{\ensuremath{\mathrm{Si}(#1)}}
\newcommand{\B}[1]{\ensuremath{B_{#1}}}
\newcommand{\Mpc}{\ensuremath{\,\mathrm{Mpc}}\xspace}

\definecolor{darkgreen}{rgb}{0.0, 0.6, 0.2}

\allowdisplaybreaks[1]

\begin{document}

\title{\textbf{\boldmath \Large \centering Living on the Fermi Edge: \\[0.2cm] On Baryon Transport and Fermi Condensation \unboldmath}}
%

\author{Andreas Trautner}
\email[]{trautner@mpi-hd.mpg.de}
\affiliation{\vspace{0.2cm}Max-Planck-Institut f\"ur Kernphysik, Saupfercheckweg 1, 69117 Heidelberg, Germany}


\begin{abstract}
The transfer function of the baryon power spectrum from redshift $z\approx 1100$ to today has recently been, for the first time, 
determined from data by Pardo and Spergel. We observe a remarkable coincidence between this function and the transport function 
of a cold ideal Fermi gas at different redshifts.
Guided by this, we unveil an infinite set of critical temperatures of the relativistic ideal Fermi gas
which depend on a very finely quantized long-distance cutoff.
The sound horizon scale of Baryon Acoustic Oscillations (BAO) seems to set such a cutoff,
which dials a critical temperature that is subsequently reached during redshift. 
At the  critical point the Fermi gas becomes scale invariant and may condense to subsequently 
undergo gravitational collapse, seeding small scale structure.
We mention some profound implications including the apparent quantization of Fermi momentum conjugate to the cutoff and the 
corresponding ``gapping'' of temperature.
\end{abstract}
\maketitle

Despite the observationally inferred presence of Dark Matter (DM) ranging from the largest scales in the observable universe down to sub-galactic scales, 
nothing is known about its corpuscular nature. Hence, the cold dark matter paradigm of the cosmological standard model $\Lambda$CDM needs to be further scrutinized in as many ways as possible, while keeping an open mind about clues inferable on the possible particle nature of DM itself.
A crucial test of DM, firstly suggested by~\cite{McGaugh:2003qw,Dodelson:2011qv}, is to track the effect of DM on baryons at large scales throughout the evolution of the universe,
captured in the so-called transport function $\Tb2$ of baryonic density perturbations. 
Crucially, this test does not require the assumption of $\Lambda$CDM 
or any other specific cosmology. Recently, Pardo and Spergel (PS) firstly extracted $\Tb2$ from measured data, 
and stressed that any theory of DM must adequately explain both its shape and normalization~\cite{Pardo:2020epc}.
While the transport function determined by PS reproduces the expectation derived under the assumption of $\Lambda$CDM reasonably well, 
the data displays a much higher level of regularity than provided by $\Lambda$CDM.
As we report in this letter, the baryonic transport function is closely matched 
(in fact, much closer than the inferred transport function in $\Lambda$CDM) 
by the red-shift transport function of a cold ideal Fermi gas. 

In order to declare this coincidence to be more than just a 
mathematical curiosity requires a full cosmological model that can be tested against the entirety of cosmological data.
The model that we are led to by the coincidence of transport functions consists of the SM amended by an effectively decoupled~\footnote{%
Effectively decoupled here means decoupled from gauge forces besides gravity and potentially effects of electro-weak interactions or small Yukawa couplings.} 
fermionic species with chemical potential $\mu$ larger than its temperature $T$, i.e.\ with a degenerate spectrum.
This could be sterile neutrinos or other new fermions and the corresponding extension of the SM Lagrangian density is straightforward.
As is well known, such a decoupled extension of the SM is easily compatible with all observational constraints if the corresponding fermions 
are either: $(i)$ heavy enough and do not contribute more to the matter density than reserved for DM, 
or $(ii)$ light and their energy density is less than the indirect bound imposable on the effective number 
of decoupled relativistic species $N_{\mathrm{eff}}$~\cite{Chen:2015dka}.

For the massive case $(i)$ ($m\gtrsim 1\,\mathrm{keV}$) it is well-known that the new fermions can be good DM candidates~\cite{Dodelson:1993je,Shi:1998km} that satisfy 
all known constraints~\cite{Drewes:2016upu}. By contrast, in the light or massless case $(ii)$ it would have to be shown that other successful 
predictions of $\Lambda$CDM, such as the matter power spectrum, temperature fluctuations of the Cosmic Microwave Background (CMB)
as well DM phenomenology on galactic scales, can be consistently explained \textit{if} the new fermions indeed are assumed to explain \textit{all} of the DM. 
The light scenario may be harder to exclude than naively expected because the exclusion of ``hot'' dark matter predominantly arises from 
structure formation which heavily relies on the use of simulations~\cite{Bond:1980ha,White:1983fcs} that are expected to be substantially altered 
by taking into account the non-trivial transport function of the DM candidate arising from its degenerate Fermi-Dirac spectrum. 
Effects on the CMB spectral fluctuations are harder to accommodate as the time of matter radiation equality would have to be altered
but earlier studies on hot and self-interacting DM~\cite{Raffelt:1987ah, Atrio-Barandela:1996suw,Hannestad:2000gt} indicate that
this could be a legitimate possibility. Finally, DM observations on galactic scales could be explained if the fermions condense to scalars 
for which the Tremaine-Gunn bound does not apply~\cite{Tremaine:1979we} or if gravity is modified for small accelerations.

For reasons of mathematical tractability we will in this work entirely focus on the case of an ultra-relativisitic Fermi gas, where $m\ll \kF$ and hence $\mu\approx\kF$ with $\kF$ being the Fermi momentum. The non-relativistic case with \mbox{$m\gg\kF$}, hence $m\lesssim\mu$, as well as a detailed investigation of CMB spectral fluctuations in the light case, are reserved for future work.

The paper is organized as follows. We proceed by giving details on the baryonic transport function and how it is matched by
the transport function of an ideal relativistic Fermi gas. Then, we investigate in detail how the transport function of the Fermi
gas comes about. Finally we discuss how the initially light fermions might reproduce the observation of all DM despite being relativistic through recombination, 
and give further comments on our findings.

\smallskip 
Given a primordial spectrum of perturbations $P_\phi(k)$ the 
power spectrum at later stages is related by a transfer function
as $P(k)\propto T^2(k)P_\phi(k)$. In analogy with this, 
a transfer function can be defined that describes
the evolution of Baryon density correlations from redshift $z=1100$ to
redshift $z\sim0$,
\begin{equation}\label{eq:T2b}
 \Tb2~\equiv~\frac{P_{bb}(k,z\sim0)}{P_{bb}(k,z=1100)}\;.
\end{equation}
To determine $\Tb2$ firstly, PS extracted the baryon power spectrum at redshift $z=1100$ from
CMB $EE$-mode polarization data \cite{Aghanim:2019ame},
and at redshift $z=0.38$ from the galaxy-galaxy power spectrum determined by surveys of BAO~\cite{Beutler:2015tla}.
\begin{figure}
\includegraphics[width=1\linewidth]{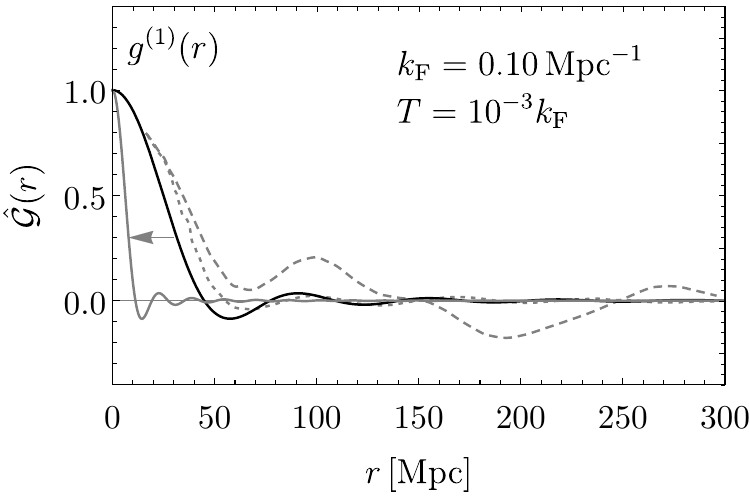}
\caption{\label{fig:G}
Our result for $g^{(1)}(r)$ (solid black) as obtained from a fit to 
$\Tb2$ taken from~\cite{Pardo:2020epc}.
The gray dashed lines show the initial estimates of \cite{Pardo:2020epc}. 
We also illustrate how the function $g^{(1)}(r)$ changes for larger values of 
$\kF$ in solid gray.}
\end{figure}

More intuitively than~\eqref{eq:T2b}, one may look at the corresponding Hankel transform, namely the position space Green's function
\begin{equation}\label{eq:Hankelxk}
 \Gb~=~G_0\int\displaylimits_0^\infty\,dk\,\frac{k^2}{2\pi^2}\,\Tb{}\,j_0(kr)\;.
\end{equation}
Throughout, $j_\alpha(x)$ denote spherical Bessel functions of the first kind.
The normalization $G_0$ here is set arbitrary, as it has not yet been determined 
from the data (naturally, it would be the density).
Under assumptions clearly formulated in PS, $\Gb$ shows the response function
that any modified gravity theory of DM must have in order to explain the evolution of baryons on large scales.  
Crucially, $\Gb$ changes sign at a scale closely related to the physical BAO scale,
implying that any ``modified gravity'' theory would have to have this scale imprinted~\cite{Pardo:2020epc}.

\begin{figure}
\includegraphics[width=1\linewidth]{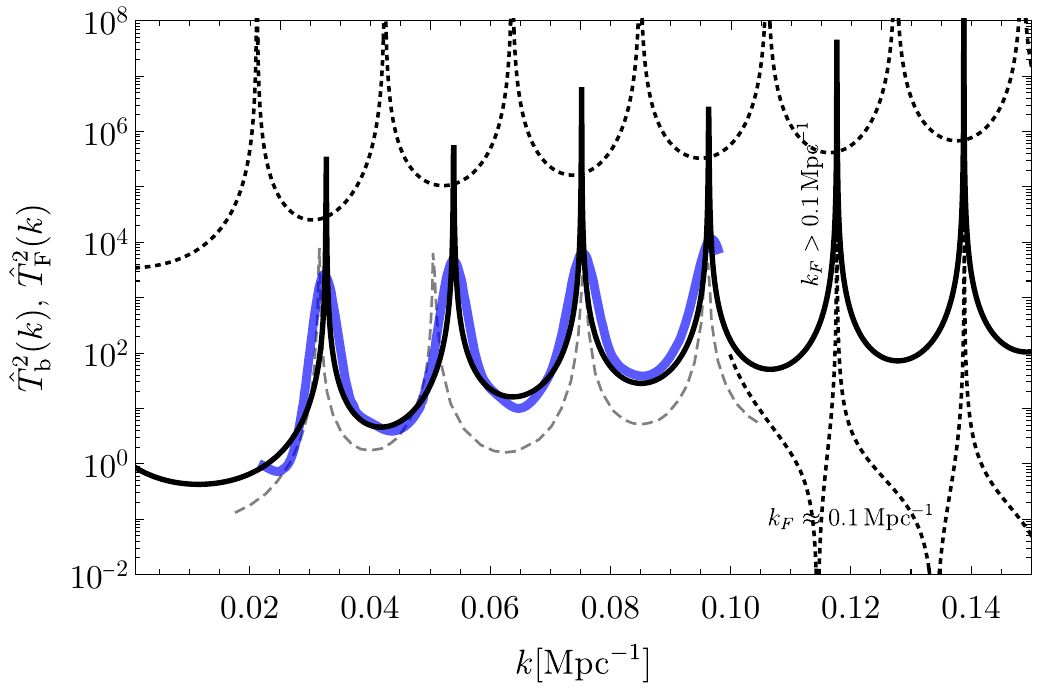}
\caption{\label{fig:T}
The measured baryon transport function \Tb2 of the large scale structure of the Universe is shown in blue~\cite{Pardo:2020epc}.
The solid black curve shows the momentum space power spectrum transport function \TF2 of the ideal Fermi gas (in first order Sommerfeld approximation) 
for representative parameters~(see Eq.~\eqref{eq:parameters}).
The dashed gray line shows the expectation for \Tb2 in the standard $\Lambda$CDM paradigm computed with CAMB~\cite{Lewis:1999bs} (extracted from~\cite{Pardo:2020epc}).
The dotted black curves show the transport function for vanishing phase shift $\kU$ (vertically offset) 
or a small value of $\kF$, see text for details.}
\end{figure}

The purpose of this memo is to point out that a well-fitting template to $\Gb$ and \Tb2, see~\Figref{fig:G} and~\ref{fig:T}, 
is given by the single-particle correlation (i.e.\ auto-correlation) function of an ideal Fermi gas (see e.g.\ \cite{Schwabl:1997gf})%
\begin{equation}
 g^{(1)}(\vec{r}):=\Braket{ \Phi_0 | \Psi(\vec{r}) \Psi^\dagger(0)| \Phi_0 }\;.
\end{equation}
The expectation value here is taken in the background of fermions, e.g.\ at $T=0$:
\begin{equation}
\ket{\Phi_0} = \prod_{{|\vec{k}|\leq \kF},\sigma} a^\dagger_{\vec{k},\sigma}\ket{0}\;,
\end{equation}
with momenta $k$ smaller than the Fermi momentum $\kF$ ($k_\mathrm{B}=\hbar=c=1$) and $\sigma$ running over spin d.o.f.'s. 
We stress that all expressions used in this work are fully relativistic.
At finite temperature $T\geq0$ and chemical potential $\mu\geq m$ we can compute $g^{(1)}(\vec{r})=g^{(1)}(r)$ from the integral%
\begin{equation}\label{eq:integral}
 g^{(1)}(r) = \frac{g}{2\pi^2}\int\displaylimits_0^\infty \frac{dp\, p^2}{\mathrm{e}^{(E-\mu)/T}+1} \frac{\sin(p\,r)}{p\,r}\;,
\end{equation}
where $g$ counts the number of spin d.o.f.'s. 
To leading order in Sommerfeld expansion we obtain
\begin{widetext}
\begin{equation}\label{eq:g_Sommerfeld}
g^{(1)}(r) = 
3n \left\{  \frac{j_1(\kF r)}{\kF r} 
+\frac{\pi^2\,T^2}{6\,\kF^2}\left[\frac{\mu^2\,r}{\kF} j_{-1}(\kF r) + j_0(\kF r)\right] \right\}
+\mathcal{O}(T^4/\kF^4)
+\mathcal{O}(\mathrm{e}^{-\mu/T})\;.
\end{equation}
\end{widetext}
\widowpenalty10000\clubpenalty10000
Here, $n$ is the zero-temperature density,
\begin{equation}\label{eq:n0}
 n=\frac{g}{6\pi^2}\left(\mu^2-m^2\right)^{3/2}\equiv\frac{g}{6\pi^2}\kF^3\;,
\end{equation}
and we introduce the exact identity $\mu^2\equiv m^2+\kF^2$ to eliminate the mass throughout.
There could be two regions of interest here, $\kF\gg m$ as well as $\kF\ll m$. 
We stress that the Sommerfeld expansion is not valid in the latter region because the 
integrand of \eqref{eq:integral} is discontinuous close to the mass threshold. 
A different expansion exists in this region~\cite{Trautner:2016ias},
but performing this for \eqref{eq:integral} is a 
challenging computation, beyond the scope of this work. 
Presently, we focus entirely on 
the region $\kF\gg m$, where $\mu\approx\kF$ to good approximation.

We Hankel transform $g^{(1)}(r)$ (inverse to \eqref{eq:Hankelxk}) to arrive at the momentum space power spectrum,
\begin{equation}\label{eq:Hankelkx}
\Tk=\int\displaylimits_\lambda^\Lambda dr\,4\pi\,r^2\,g^{(1)}(r)\,j_0(k r)\;.
\end{equation}
Here we have included short and long-distance cutoffs $\lambda$ and $\Lambda$ whose meanings we see momentarily. 
$\Tk$ can be computed analytically, be there cutoffs or not, and we state analytic expressions in Eqs.~\eqref{eq:Tkintegral} and \eqref{eq:TkExpl}.
If we set the cutoffs to their maximally allowed range $(\lambda\rightarrow0,\Lambda\rightarrow\infty$), the power spectrum 
is \textit{practically} a box~\footnote{\label{fot:FDbox}
A more accurate expression in the limit $\Lambda\rightarrow\infty$ includes Dirac-$\delta$ distributions and corrects this expression by
$\Delta\Tk=g\,T^2\pi^2/(6\kF^2)\left[\kF\delta(k-\kF)+\mu^2\delta'(k-\kF)\right]$.
In fact, in the limit $\Lambda\rightarrow\infty$ one can compute $\Tk$ to all orders in $T$ by using 
the integral representation in Eq.~(\ref{eq:Tkintegral}) with the result
$\Tk^{\infty}_{0}=g\left\{\mathrm{exp}\left[\left(\sqrt{k^2+m^2}-\mu\right)/T\right]+1\right\}^{-1}$.
}
\begin{equation}\label{eq:box}
\Tk^{\infty}_{0}=g\,\Theta(\kF-k)\;,
\end{equation}
with $\Theta$ being the Heaviside function. 
This is the usual box of the Fermi-Dirac distribution, see~\cite{Note2} for finite-$T$ corrections.
We emphasize that a power spectrum of the Fermi-Dirac shape can only be obtained 
under the tacit assumption of being able to probe the fermions at arbitrary length.
More realistically, there is a maximal possible length at which the Fermi gas can be probed
implying that a physical long distance cutoff should be introduced. In a laboratory setup with sufficiently long measurement times this 
would correspond to the size of the apparatus or trapping potential, while in a cosmological situation the cutoff is bounded 
from above by the respective causal horizon.

We stress that \Tk, absolute squared and normalized, corresponds to the power spectral density which can be assigned a 
spectral entropy, i.e.\ this curve has an information-theoretic meaning. 
The box corresponds to white noise with wave numbers $k\leq\kF$.
Hence, the cutoffs are crucial to obtain a response to $g^{(1)}(r)$ with finite spatial resolution, 
which leads to more interesting results for~$\Tk$ as we will discuss in detail below. 

To turn the power spectrum into our desired cosmic conveyor belt we have to evaluate it, relatively, at different 
cosmological redshifts $z$. In this way we obtain our fitting template for
the transport function of the Fermi gas
\begin{equation}\label{eq:transport}
  \TF2 = N\; \frac{\mathcal{T}(k,\kF,\mu,T,\lambda,\Lambda)^2}{\mathcal{T}(k,z\,\kF,z\,\mu,z\,T,\lambda,\Lambda)^2}\;,
\end{equation}
including an arbitrary normalization $N$ that can currently not be fixed from the data.
We have very explicitly spelled out all parameter dependencies of $\Tk$ here, to be clear which quantities transform
under redshift, or in other words, scaling transformations.
The cutoffs do not red-shift because they correspond to the fixed resolution at the high $z$ probe scale, while $k$ does not shift
because it is our ruler at the low $z$ probe scale. 
In addition to the parameters listed, we find it necessary to introduce a phase shift $\hat{T}^2_{\mathrm{F}}(k+\kU)$
that we also include as a fit parameter.

We use our template for the transport function to perform a 
simple MCMC fit to the PS measurement~\cite{Pardo:2020epc}.
One result is shown as the solid black line in \Figref{fig:T},
next to the data extracted from PS in solid blue.
The following parameters play straightforward roles in determining a good fit point: 
$(i)$ $\kU$ is essentially determined by aligning the first peak to data in horizontal direction.
$(ii)$ $\Lambda$ sets the period of oscillations, i.e.\ it is tightly fixed by the observed peak-to-peak distance.
$(iii)$ $\kF\approx\mu$ sets the size of the relevant ``box'' and, hence, determines the number of complete resonant 
peaks of \TF2 which are located \textit{in} the Fermi sphere.
Next, there are some parameters that do not play so relevant roles after all:
$(iv)$ Everything is largely insensitive to the precise value of the UV cutoff $\lambda$, as it should be,
and so we fix it to $0.5\,\mathrm{Mpc}$.
$(v)$ Varying $0\leq m\ll\kF$ within the validity of Sommerfeld has no effect and we set $m=0$ (hence $\kF=\mu$) 
for the sake of the fit.
Finally, two parameters that play very subtle roles in the fit are redshift and temperature, more details
below. We fix $z=1100$ in the ultimate 
fit to comply with~\eqref{eq:T2b}, while noting that a redshift interval of about $\Delta z\approx10$ is enough
to create the peaks required. We stress that $T/\mu$ must be finite 
to make the fit work. The curve shown in \Figref{fig:T} is obtained for 
\begin{align}\notag
 \kU &= -0.0115\,\mathrm{Mpc^{-1}},& 
 \Lambda &= 148\,\mathrm{Mpc},& \\ \label{eq:parameters}
 T/\kF &= 1\times10^{-3},&
 \kF &= 0.45\,\mathrm{Mpc}^{-1}.&
\end{align}
$\Lambda$ here clearly corresponds to the scale of the BAO sound horizon, but there could be much more to it: 
The fit shows that the best fit values are obtained with \textit{discretized} values for \kF.
The Fermi momentum seems to be quantized in units of
\begin{equation}\label{eq:kFquantization}
\kF=\kU+\nu\,\Delta\kF \quad\text{with}\quad \Delta\kF\approx\frac{2\pi}{\Lambda},\,\nu\in\mathbbm{N}.
\end{equation}
In these units, the best fit value for $\kU$ corresponds to a $-\pi/2$ phase shift. The vertically offset dashed curve 
in \Figref{fig:T} is obtained for $\kU=0$. A minimum of $\nu=3$ (i.e.\ $\kF\approx0.1\Mpc^{-1}$) is required for \kF to explain the observed data,
but it could also be much larger. We show the possible extrapolations to larger $k$ depending on the size of \kF in \Figref{fig:T}.

Let us also show the resulting $g^{(1)}(r)$, see black line in \Figref{fig:G}.
Note that PS had trouble in extracting this function
from the data, as performing the integral \eqref{eq:Hankelxk} 
depends on the extrapolation of \Tb2 to momenta outside
of the observed region. We do not have this problem here
since we started from $g^{(1)}(r)$ and performed the inverse transformation
to obtain the power spectrum. Hence, $g^{(1)}(r)$ is fixed by the fit to \eqref{eq:T2b},
besides the discrete choice of $\nu$. 
The resulting function is shown as the solid black line in \Figref{fig:G},
for minimum allowed $\kF$, together with the initial estimates of PS.
For larger $\kF$, $g^{(1)}(r)$ slides as indicated by the arrow and gray line in \Figref{fig:G}.

\medskip
So what are we looking at here? So far we fitted the scale-transport function
of this innocent Fermi gas to the observed baryonic transport function of the Universe.
If one has to do with the other, baryons have to interact with this momentum space lattice.
One possible scenario could be that baryons directly scatter off the Fermi gas 
with a cross section $\sigma$ and mean free path $\lambda_{\mathrm{mfp}}=(n\sigma)^{-1}$.
In this case, the fermions would act as a low-pass filter for momentum, 
as low-momentum modes may not be excited in the Fermi gas.
Having the baryons scatter at least once in a $150\Mpc$ would require
a cross section 
\begin{equation}
 \sigma\approx 1\times10^{-39}\,\mathrm{cm}^2\left(\frac{1\,\eV}{\kF}\right)^3\frac{1}{g}\;.
\end{equation}
Compared to an electro-weak cross section of momentum transfer $\kF$ this would require 
$\kF g^{1/5}\approx2\,\keV$, implying an energy density
in the Fermi gas that would overclose the universe.

Alternatively, recall that baryon transport is usually ascribed to DM, 
implying that long range gravitational-strength interactions seem to suffice in order to imprint
the transport functions of the fermions onto the baryons.
Supposing that our fermions would contribute an energy density akin to that of all the DM, 
an estimate of the required Fermi momentum at recombination is
\begin{equation}\label{eq:kFBAO}
k_{\mathrm{F,*}}=\left(\frac{8\,\pi^2}{g}\rho_{\mathrm{DM},0}\,z^{3}_*\right)^{1/4}\approx \frac{1.0\,\mathrm{eV}}{g^{1/4}}\;.
\end{equation}
This is not an incredibly large chemical potential.
However, if stored in standard model neutrinos, a chemical potential of this size would violate the BBN bound on 
neutrino degeneracy~\cite{Cuoco:2003cu, Serpico:2005bc, Mangano:2011ip, Oldengott:2017tzj}. 
The chemical potential could also be stored in a non-thermal 
background of right-chiral neutrinos~\cite{Chen:2015dka}, in which case the
maximal allowed energy density during recombination expressed in $\Delta N_{\mathrm{eff}}$ \cite{Aghanim:2018eyx},
results in a constraint $k_{\mathrm{F},*}\lesssim1.85\,T_{\gamma,*} (\Delta N_{\mathrm{eff}}/g)^{1/4}$, 
only in mild tension with \eqref{eq:kFBAO}.
While these bounds might easily be avoided in more elaborate models, another possibility is that the fermions 
have a mass and turn non-relativistic in the vicinity of recombination.
In fact, the required $k_{\mathrm{F,*}}$ is awkwardly close to the sum of the observed neutrino 
masses.
We remind the reader though, that computing \eqref{eq:integral} in a region where $\kF\sim m$ does 
require more care. At this stage, one may argue that these fermions can impossibly be 
the DM we observe on galactic scales given the seminal bound by Tremaine and Gunn \cite{Tremaine:1979we}.
Note that in the natural quantization imposed upon us by $\Lambda$, 
this value of $k_{\mathrm{F},*}$ would correspond to a large number of $\nu\approx 3.7\times10^{30}$ nodes in the Fermi sphere.

\medskip 

\begin{figure*}[!t!]
\begin{center}
\includegraphics[width=0.45\linewidth]{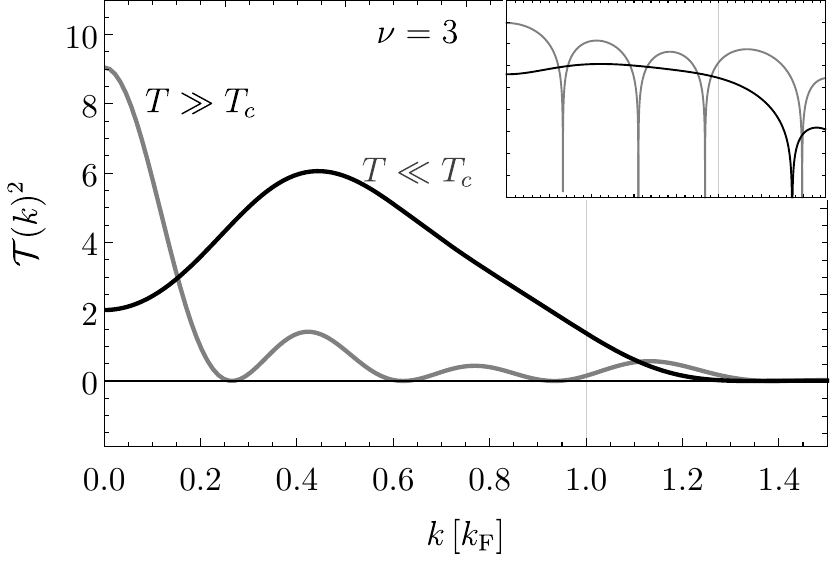}
\hspace{1cm}
\includegraphics[width=0.45\linewidth]{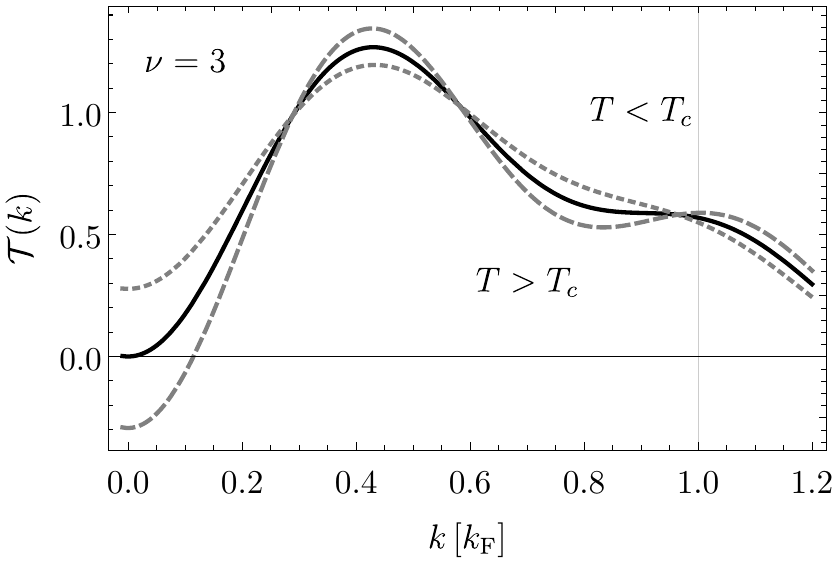}
\end{center}
\caption{\label{fig:PTI}
The power spectrum behavior during the phase transition for temperatures $T[\mu]$ ($\kF$ fixed to $\kF=\mu$ for the plots).
LHS: values for $T[\mu]$ are $0.09$ and $0.3$ and the curves have been rescaled by factors of $5$ and $10^{-1}$, respectively, to fit the plot.
The inset shows the same plot on a log-scale to better visualize how the transport function (cf.~\Figref{fig:T}) comes about.
RHS: Behavior around the lowest-temperature critical point $T^2=T^2_{c,0}$ for  $T/\mu=0.105, 0.125$ and $0.115158$ $(\nu=3)$. 
While we display the behavior of the power spectrum as a function of actual temperature here,
we stress that it undergoes the very same events also as a function of redshift.}
\end{figure*}
\begin{figure}[b]
\includegraphics[width=1\linewidth]{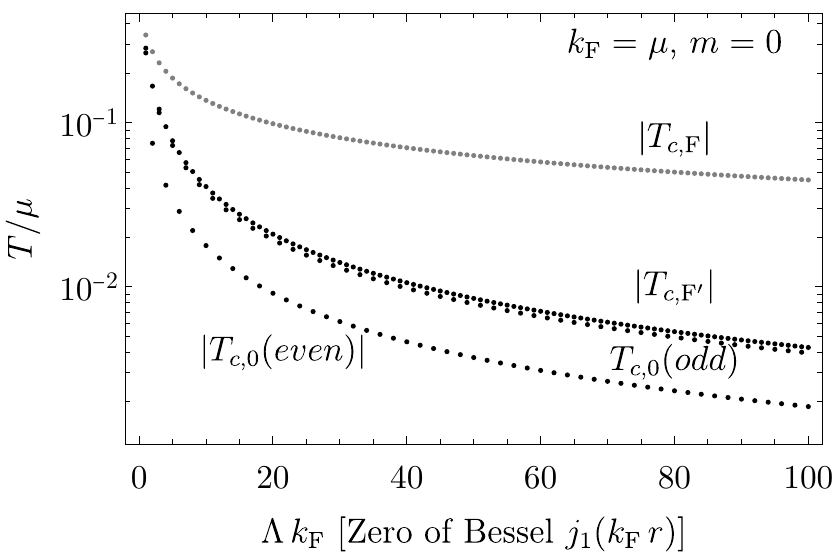}
\caption{\label{fig:TCrit}
Critical temperatures as function of the ``Bessel cutoff'' $\Lambda$. 
Exact expressions are stated in \eqref{eq:Tcrit0}, \eqref{eq:TcritFermi} and~\eqref{eq:TcritDFermi}.}
\end{figure}
We leave them there for now and pick up the discussion on the required sizes of $z$ and $T$. 
We noted from our fit that $T/\mu>0$ is required, and that $\Delta z\approx 10$ was enough to make peaks 
appear in the transport function, \Figref{fig:T}. In fact, we noted that the \textit{absolute} value of temperature surpassed 
in the redshift sweep seemed to play a role. How can that be, given that $T/\mu$ is scale invariant?
Note that in the case of an infinite cutoff $\Lambda$, the resulting box power 
spectrum \eqref{eq:box} is \textit{almost} scale invariant.
While the height of the box is scale invariant even for finite $T$, $\mu$ and $\kF$,
it is our \textit{probe scale} ruler $k$ that leads to an \textit{explicit} breaking 
of scale invariance in the theta function. 
Moving on to the more general case including cutoffs, we note that
also the cutoffs break scale invariance explicitly. Yet in a very subtle manner:
Given the quantization indicated by momentum space matter oscillations,
see \eqref{eq:kFquantization}, 
it seems to make sense to fix the long-distance cutoff in \eqref{eq:Hankelkx} to a 
definite zero of the integrand. 
For $T=0$ those are given by the $\nu$-th zeroes 
of the Bessel function $j_{1}(\kF x)$, in the following called \B{\nu}.
In fact, note that the best fit $\Lambda$ itself 
corresponds almost exactly to (at least) the fourth zero of the Bessel function 
(the zero crossing at $\Lambda$ has also been stressed by PS).
For finite $T$ the zeros of $g^{(1)}(r)$ get slightly misaligned 
with $\B{\nu}$. Nonetheless, $\B{\nu}$ seem to provide exquisite choices 
of long distance cutoffs. 

Taking our consideration of the box above as motivation, we thrive now to find temperatures
at which the resulting power spectrum $\Tk$ \textit{might} become scale invariant. 
Given a quantization of $\Lambda$ in units of $\B{\nu}/\kF$, as suggested by the data
\footnote{%
Note that our fit to the data actually seems to indicate a quantization of $\kF\times\Lambda$ in units of $\B{2\nu}$ or $\B{2\nu+1}$. 
Nonetheless we press on with the more general case here as not to loose any information on the way.
},
and setting $\lambda=0$ we find that there are two special points in the resulting 
power spectrum, namely $k=0$ and $k=\kF$. Requiring that the spectrum vanishes at these points 
allows us to implicitly define two critical temperatures
\begin{equation}
\mathcal{T}(0,T_{c,0})\stackrel{!}{=}0\;, \quad \text{and}\quad \mathcal{T}(\kF,T_{c,\mathrm{F}})\stackrel{!}{=}0\;,
\end{equation}
while requiring a vanishing derivative yields a third,
\begin{equation}
\frac{d}{dk}\mathcal{T}(k,T_{c,F'})|_{k=\kF}\stackrel{!}{=}0\;.
\end{equation}
The resulting temperatures are functions of $\mu$ and $\kF$, as well as the cutoff, parametrized as \B{\nu}.
We display these temperatures in \Figref{fig:TCrit} and state exact expressions for them in \eqref{eq:Tcrit0}, \eqref{eq:TcritFermi} and \eqref{eq:TcritDFermi}. 
The absolute values of all these temperatures are, to our understanding, 
in a perfectly valid region of the Sommerfeld 
expansion which is trustworthy for 
\begin{equation}\label{eq:noTc}
 T^2\ll\frac{2\,\kF^4}{\pi^2(\kF^2+\mu^2)}= \frac{\mu^2}{\pi^2}\;,
\end{equation}
where the last equality holds in case $m=0$.
Nontheless, note that only the critical temperatures \mbox{$T_{c,0}(2\nu+1)$} are positive.
$T_{c,0}(2\nu)$, $T^2_{c,\mathrm{F}}(\nu)$, and $T^2_{c,\mathrm{F'}}(\nu)$ are negative for all $\nu$ (mind the squares).

At this point we give a disclaimer, stating that the investigation of the tantalizing phase transition happening around these 
critical points will undoubtedly require much more scrutiny and care than what we can deliver in this short paper. 
Thus, everything that we have to say must necessarily sound premature and speculative. 
We will not further touch regions with imaginary critical temperatures in this paper but we note that they are special.
One may without problem rotate the temperature to imaginary values, while the power spectrum stays a real function.
Rotations of this kind affect the exponential correction to the Sommerfeld expansion \eqref{eq:g_Sommerfeld},
and therefore might, together with imaginary values of the chemical potential, 
transfer a density of particles from one sector to another. 

We now focus on the real squared temperatures in $T^2_{c,0}(\nu)$ because the reader may more comfortably be
convinced that real temperatures exist. 
At all critical temperatures $T^2_{c,0}(\nu)$ the power spectrum \textit{is} scale invariant (in the sense of self-similar) 
under the remaining redshift capabilities of $\mu$ and $\kF$.
For far above and below the critical temperature(s), we show the according power spectrum for the 
example choice of a cutoff $\nu=3$ in \Figref{fig:PTI} (left). The behavior around the critical 
temperature $T^2_{c,0}(2\nu+1)$ is highlighted in \Figref{fig:PTI} (right).
We observe that $T^2_{c,0}(\nu)$ is the endpoint of a dramatic series of events, turning 
the initial state of the power spectrum (gray line in \Figref{fig:PTI}, left) into a final state (black line in \Figref{fig:PTI}, left).
In the process, several nodes, minima, maxima and turning points (in particular, their corresponding information) 
are ejected from the Fermi sphere. The process shown in the right of \Figref{fig:PTI}
(corresponding to the behavior around the only \textit{real} critical temperature in our basis) 
is merely the swirling-off of the last extremum. 
Note that all modes located outside of the Fermi sphere appear to be crossed by some of the escaping modes; 
a process during which most likely they get entangled. 

\medskip

\begin{figure}[h]
\includegraphics[width=1.0\linewidth]{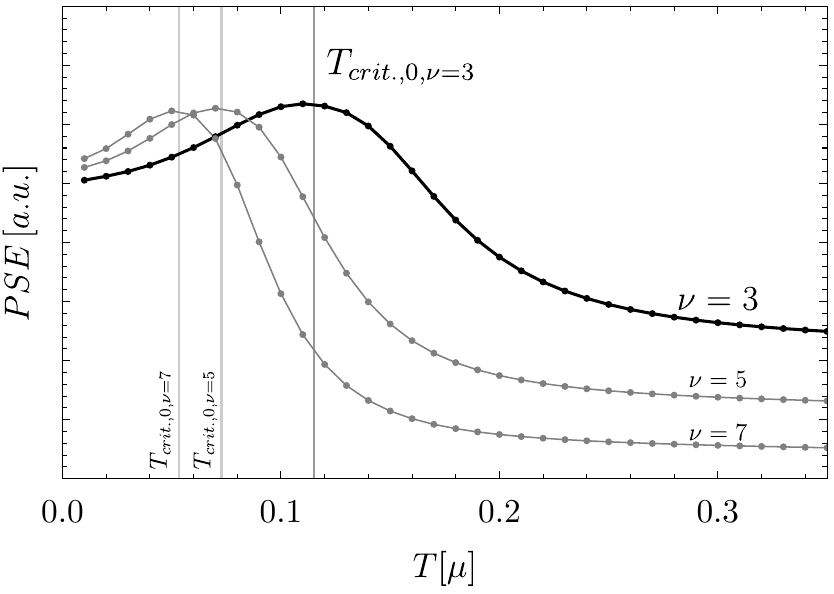}\hfill
\caption{\label{fig:PSE}
Power spectral entropy (PSE) of the power spectrum $\Tk$ as a function of temperature for different choices of the long-distance cutoff (parametrized as the $\nu$-th zero
$B_\nu$ of the spherical Bessel $j$-functions).
The horizontal lines mark the independently computed critical temperatures $T^2_{c,0}(\nu)$ to show that they line up with stationary points of the PSE.
}
\end{figure}

To corroborate the information-theoretic nature of this phase transition we compute the spectral entropy
of the power spectrum, given by a generalization of Shannon's discrete entropy~\cite{Shannon:1948zz} 
to continuous probability distributions~\cite{Jaynes:1963}.
The power spectral density corresponds to a probability density 
\begin{equation}
p(k)\,dk := \frac{|\Tk|^2\,dk}{\int\displaylimits_0^\infty\,|\mathcal{T}(k')|^2\,dk'}\;,
\end{equation}
that allows us to compute the power spectral entropy
\begin{equation}
 PSE(\mu,T)=-\int_0^\infty\,dk\,p(k)\,\ln\left(\frac{p(k)}{f(k)}\right)+const.
\end{equation}
Here, $f(k)$ is a necessary normalization for continuous probability distributions~\cite{Jaynes:1963} 
that we take to be the usual Fermi-Dirac density of states $f(k)=dn/dk$.
Taking $k$ in units of $\kF$, $\mu\sim\kF$ and $T$ in units of $\mu$ all of these integrals can be computed 
numerically and converge quickly. We display the resulting power spectral entropy in \Figref{fig:PSE} for the case $\nu=3$,
and highlight that it becomes stationary at (or very close to) the critical temperature.

\medskip 

\begin{figure}[t]
\includegraphics[width=1\linewidth]{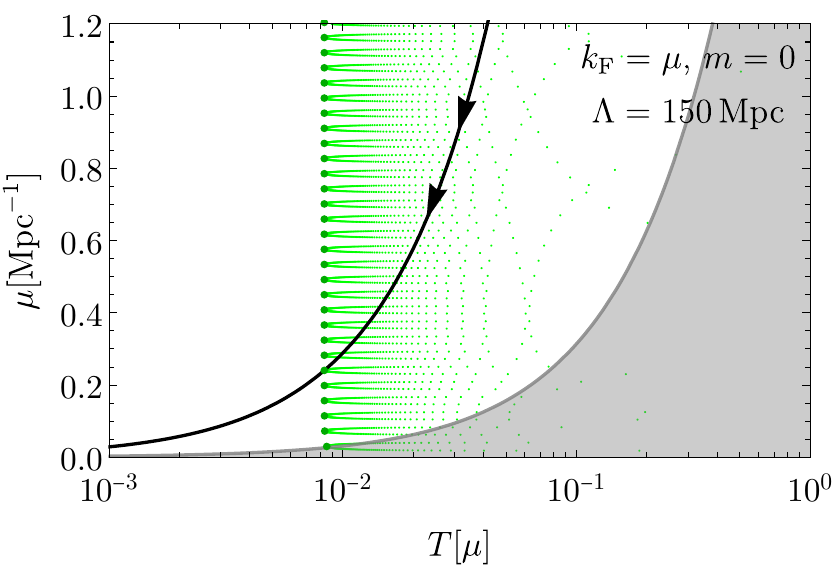}
\caption{\label{fig:muT}
Critical temperatures (green) of the ideal relativistic Fermi gas shown in the $T$-$\mu$ plane for the example choice $\Lambda=150\,\mathrm{Mpc}$ 
(the discreteness in the horizontal direction arises because we sample $\mu$ with a finite resolution, the discreteness in the vertical
direction is genuine).
The thicker green dots show the critical temperatures obtained for the special values $\kF=\mu=B_\nu/\Lambda$.
The shaded gray region is where the Sommerfeld expansion is not valid. The black line shows the typical evolution of 
$T$ and $\mu$ under cosmological redshift indicated by the arrows (ignoring anything that might happen at the critical points).}
\end{figure}

\begin{figure}[t]
\includegraphics[width=1\linewidth]{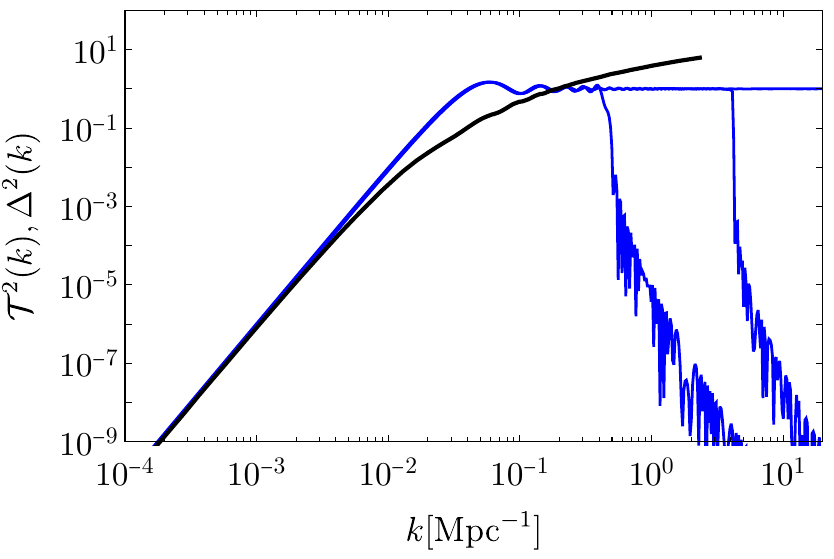}
\caption{\label{fig:MPS}
Power spectrum of an ideal Fermi gas at the $T_{c,0}$ critical point for excitation numbers $\nu=11,100,10^{30}$ (overlapping, in blue), the lines break at their respective Fermi edge. 
Nothing enters this plot besides the BAO scale which sets the horizontal offset of the scale invariant (blue) curves by providing a (cutoff-)ruler of the size $150\,\mathrm{Mpc}$.
We also show the Universes' matter power spectrum extracted from~\cite{Akrami:2018vks} (using $\Delta^2(k)\equiv k^3(2\pi^2)^{-1}P(k)$ and $h=0.7$).
We have not touched the normalization of the curves.
}
\end{figure}

Let us come back to the actual situation of baryon transport, and reset the cutoff $\Lambda$ as well as $\kF$ to be free and independent parameters. 
In \Figref{fig:muT} we show the critical temperatures in the $T-\mu$ plane for an example cutoff of $\Lambda=150\,\mathrm{Mpc}$ next to the validity region of the Sommerfeld
expansion, \Eqref{eq:noTc}, and an arbitrary example for the usual evolution of $T$ and $\mu$ under redshifts.
Clearly, it is not necessary to tailor $\Lambda$ to a Bessel cutoff to obtain a critical temperature akin to $T_{c,0}$, where scale invariance is restored 
(the critical temperatures obtained for Bessel cutoffs provide lower bounds for critical temperatures obtained with all other cutoffs).
Crucially, note that the transport of modes in the power spectrum as a function of redshift proceeds very similar to the transport 
as a function of temperature, best visualized from the inset in \Figref{fig:PTI} (left).
The illustrated transport of modes \textit{is} the behavior displayed in the observed structure-formation transport function of the Universe, \Figref{fig:T}. 
Given $T/\mu$ of our potentially structure-forming Fermi gas and the overall appearance of its transport function, 
we conclude that the fermions that might be responsible for the large scale structure of the Universe have already undergone this phase 
transition; i.e.\ data that tells us, we sit below the critical point.
At the critical temperature the Fermi gas becomes scale invariant, and observation indicates that the fermions
get stuck at this symmetry enhanced point:
So far we had looked at the transport from redshifts of $z\approx1100$ down to today,
fitted to the observed baryon transport function.
Having the transport anchored at today, as in Eq.~\eqref{eq:transport}, we might as well look at the 
``whole'' transport, say down from $z\approx 10^{10}$ to today, and find that it does not differ much from the 
one down from $z\approx1100$. 
On the other hand, if the power spectrum were fixed for redshifts below the critical point
this would allow us to compute the \textit{absolute} power spectrum of the fermions irrespective of further redshifts.
We show the power spectrum of the ideal fermions at the critical point in \Figref{fig:MPS}, together with today's observed baryon power spectrum. 
While the scale invariant power spectrum of the fermions gives a good leading order approximation, it does not coincide with the observed matter power spectrum
on small scales. Simulations would be necessary to see whether this situation can be improved if the fermions condense to scalars, 
which plausibly could clump in order to transfer power from larger to smaller scales in the course of the evolution of the universe.

\smallskip

\enlargethispage{0.8cm}
As of now, we cannot with certainty tell the nature of the final state after the phase transition,
but it is a logical possibility that the structure-forming fermions undergo condensation~\footnote{%
An alternative to condensation would be that the fermions simply stick around until today, which
would be no problem as their energy density would redshift away $\propto (\Delta z)^{-4}$. Nonetheless,
this seems to be the least elegant option here, as there would then still be the need for an additional 
cold DM component on small scales. See~\cite{Coleman:1974bu} for more aspects of fermion-boson 
metamorphosis beyond all conjecture.}. 
Following this hypothesis, the final state that we would be looking at from below in redshift 
appears to be of the size of the Fermi sphere in momentum space and bosonic.
Subsequent to condensation, the energy density of the condensed state would redshift $\propto (\Delta z)^{-3}$ like ordinary matter.
A logical possibility is that the fermions form quasi-particles akin to Cooper pairs that condense into a superfluid stage. \linebreak
If each two fermions make a boson and if all of the energy would be stored in the boson masses, $\rho_{\mathrm{DM},*}=n m_b/2$, 
then \Eqref{eq:kFBAO} implies an upper limit $m_b\lesssim 1.5\,\eV g^{-1/4}$. The Jeans length for a Bose condensate of such scalars (see e.g.~\cite{Khlopov:1985jw,Suarez:2017mav}) is
$\lambda_J\sim 6\times 10^{7}\,\mathrm{m}$ and the Jeans mass $M_J\sim3\times10^{8}\,\mathrm{kg}$. 
These characteristics are sufficiently close to the stability curve of Bose-Einstein condensates~\cite{Ruffini:1969qy}
suggesting such scalars would, at least initially, form fluffy Bose ``meteors'' of this size and mass
which sit dense in position space.
The non-linearity scale of the initial power spectrum, cf.\ \Figref{fig:MPS}, i.e.\ the point when the variance of density perturbations $\sigma^2\gtrsim1$, is roughly 
$k\approx0.04\,\Mpc^{-1}$ corresponding to a non-linear evolution of structure formation on scales $R\lesssim 25\Mpc$.

We emphasize that the details of this low energy phase might be as rich as the condensed superfluid phase of ${}^3\mathrm{He}$~\cite{Lee:1997zzh}, which, 
despite being observed in the laboratory, still bears many mysteries~\cite{Volovik:2003fe}.
Hence, the quasi-particles could also be much lighter than above bound with the additional energy density stored in 
coherent field oscillations, as for the axion and similar light scalar DM candidates. Also, already miniscule self-interactions among the quasi-particles 
would crucially affect their properties under gravitational collapse and, hence, might drastically affect the evolution of 
structure formation~\cite{Chavanis:2011zi}.
Altogether, hence, small scale structure formation after the phase transition crucially depends on the details of the 
low energy phase and its excitations. These states may very well behave as previously discussed candidates for
non-relativistic cold DM such that small scale structure formation down to galactic scales and below may proceed more or less ``as usual''. 
To further test this idea, it would be extremely important to have numerical $N$ body simulations 
that go beyond the standard implementations of Maxwell-Boltzmann gases in order to simulate degenerate Fermi gases 
with correct statistics and possible mixed phases of (non-)condensed fermionic quantum gases. 
In addition, to investigate (or simulate) experimentally the details of the low-energy phase would invite laboratory studies 
of analogue systems with condensates of relativistic (i.e.\ massless Weyl) fermions, which unfortunately have not been realized to our knowledge.

\smallskip

\enlargethispage{0.8cm}
It might also be instructive to look at this phase transition proceeding from a thermally fluctuating phase.
Even in scale invariant expansion, both $T$ and $\mu$ scale down with redshift. 
The scale invariant temperature $\Theta:=T/\mu$ might perform random fluctuations, 
bare any other scale with an expectation value $\langle\Theta\rangle=0$.
However, this place is doomed as the closer one gets to zero, the more likely one will fluctuate into one of the critical temperatures.
This seems to be an artfully crafted selection mechanism for a random, but steady population of the critical points.
One may want to closer investigate this mechanism to decide whether temperature is really ``gapped'' in this way at a fundamental level.
Nontheless, we emphasize that in the case discussed here, it seems that it is not temperature fluctuations triggering the phase transition
but we rather red-shift into one of the critical points.

\smallskip

There remains the nagging phase shift $\kU$ (the well converging fit result is $\kU=-0.011260(5)\,\Mpc^{-1}$).
It is tempting to gloss over it, because $\kU$ quickly becomes irrelevant in the total $\kF$ for growing occupation number
in \eqref{eq:kFquantization}.
However, the phase shift is absolutely relevant for the transport function at low $k$. 
In other words: we are observing this offset phase shift already in the ``first bin'' in $k$ such that 
adjusting $\kU$ correctly is absolutely crucial in order to fit the data.
Hence, understanding the precise value of $\kU$ might be a key check that we are correctly 
interpreting the dynamics of the structure forming phase transition.
From a physical point of view, the offset implies that all baryonic matter seems to get a little push 
relative to the initial fermions. 
Nonetheless, we can presently not compute $\kU$, and so this remains an open question.
Also, even though we think this would be very tempting, we did not succeed in relating $\kU$ to any 
of the observed dipoles in the Universe~\cite{Secrest:2020has, Siewert:2020krp}.

\smallskip

Finally, it is interesting to think about what causes the breaking of scale invariance at a distance of $150\,\Mpc$.
While it might just be the physical scale of the BAO sound horizon, 
we note that there is an accidental proliferation of scales in the vicinity of $150\,\mathrm{Mpc}$.
By chance, this also falls close to the size of our physical horizon 
in neutrinos today, as well as to the neutrino comoving travel distance until recombination~\cite{Dodelson:2009ze}.
In fact, carefully considering the arguments of the present paper, one may currently not exclude 
that the spatial cutoff $\Lambda$ itself might be quantized conjugate to $\kF$.
Coming from the ultrarelativistic regime and approaching $(\kF)_{\mathrm{min}}\gtrsim m$ this would imply that 
the fermion properties themselves lead to an upper bound on the long-distance cutoff $\Lambda\lesssim \pi \nu/m$.
Following this path of thought implies that similar phase transitions might occur every time the 
lightest species of fermions becomes non-relativistic by redshift and is forced to see the critical temperatures.
This could mean that $2m_e\sim T_{\mathrm{BBN}}$ might not be an accident.
In any case, we stress once again that our computation is not valid if any of $T$ or $\kF$ approach the mass, 
and so these speculations may only be substantiated once the full computation becomes available.
Irrespectively, we think it will be very interesting to explore the potential 
for baryogenesis in this mechanism.

\smallskip

Lastly, we point out what we think are the biggest differences of this scenario with respect to the standard cold DM paradigm. 
In $\Lambda$CDM, matter domination after $z\approx3400$ is required for structure formation on all scales.
In particular, DM needs to be non-relativistic long before CMB decoupling in order to allow DM to form early structures
that the baryons can collapse onto after recombination. 
By contrast, in the scenario hinted at here, the relativistic fermions should drag along the baryons after recoupling to explain the remarkable
coincidence of their transport functions. 
Hence, the phase transition in the Fermi gas, which causes the transport, ought to happen at a time between recombination and today.
The fact that the fermions may only red-shift as non-relativistic matter after undergoing their phase transition 
and potentially condensation implies that the turnover scale of matter-radiation equality might be delayed 
to redshifts $z_{\mathrm{turnover}}\approx1000$ (or even lower) in this scenario. 
While this may not be a problem per se, as it is still of the same order of magnitude as in $\Lambda$CDM,
it shows that a crucial test of this scenario would be to check whether or not it can accommodate 
the observed CMB spectral fluctuations. 
While this certainly has the potential to shelve the whole idea, performing such an analysis is 
beyond the scope of this short memo. In addition, the fact that the phase transition only requires a rather narrow redshift interval 
$\Delta z\sim\mathcal{O}(10)$ suggests that also the baryon transport might take much less time than in concordance cosmology, 
where it is believed to have built up rather steadily between recombination and today. 
Even though there is presently no redshift resolved measurement of the baryon transport (at least not at high redshifts), exploring the consequences 
of such a fast baryonic transport might be an interesting target for simulations of structure formation. 
Other ways to move forward and better discriminate this idea from standard $\Lambda$CDM include a more precise determination of the 
total baryonic transport function including, in particular, the inter-minimum slope, the homogeneity of peak-to-peak distances and, 
of course, the premier determination of the function on larger and smaller $k$ scales.

\smallskip

To summarize, we have pointed out that the recently firstly determined transport function of
baryons in the Universe bears remarkable coincidence with the transport function of a degenerate, 
relativistic Fermi gas. The characteristic features of the baryon transport function are reproduced by the fermions 
while they undergo a new type of information-theoretic phase transition of the power spectrum that we have firstly described here.
If both transport functions are indeed related, data seems to point to a quantization of Fermi momentum conjugate 
to a spatial cutoff, implying also a gap in the minimal possible temperatures attainable for ideal and relativistic 
degenerate Fermi gases.
To fully comprehend the new low-temperature phase of the fermions and the subsequent structure formation
on small scales will require a concerted effort of condensed matter theory, on the one hand,
and advanced numerical simulations of cosmic structure formation on the other hand.

Despite the fact that our revelations appear to be dramatic,
it seems like we would not have to abandon any of our paradigms.
We surely hope that the outlined ideas for the formation of large scale structure
as well as the hypothesis of the information-theoretic Fermi-condensation phase transition stand up 
further scrutiny. This would herald a new age of large scale cosmology in surprising 
unison with theories of condensed quantum matter.

\smallskip
I am grateful to Jonas Rezacek, Luca Amendola, Alexei Yu Smirnov, Andrei Angelescu, Evgeny Akhmedov, Christian D\"oring, Johannes Herms, Sudip Jana, 
Jeff Kuntz, Kris Pardo and David Spergel for useful discussions and comments.
I want to stress that the original Fig.~3 of PS inspired this work, and it would not have happened if Fig.~3 would only 
have appeared directly in its final (refereed) form. 

\onecolumngrid
\appendix
\section*{Appendix}\noindent
Here we state expressions for the power spectrum defined in \Eqref{eq:Hankelkx} (for simplicity with vanishing short distance cutoff $\lambda=0$).
Using the exact expression for $g^{(1)}(r)$ of \eqref{eq:integral} in the definition of \Tk, and interchanging the integrals we obtain an integral representation 
for \Tk given by
\begin{equation}\label{eq:Tkintegral}
\Tk~=~\frac{g\,\Lambda}{\pi\,k}\int\displaylimits_0^\infty \frac{dp\, p}{\mathrm{e}^{(E-\mu)/T}+1} \left\{j_0\left[\Lambda\left(p-k\right)\right]-j_0\left[\Lambda\left(p+k\right)\right]\right\}\;.
\end{equation}
To leading order in Sommerfeld expansion this evaluates to (this is consistent with first expanding $g^{(1)}(r)$ to $\mathcal{O}(T^2/\kF^2)$ as in \eqref{eq:g_Sommerfeld} 
and then performing \eqref{eq:Hankelkx})
\begin{equation}\label{eq:TkExpl}
 \begin{split}
\Tk~=~
& \frac{g}{\pi}\left\{\mathrm{Si}\left[\left(\kF-k\right)\Lambda\right]+\mathrm{Si}\left[\left(\kF+k\right)\Lambda\right] 
- \frac{2\,\sin\left(\kF\Lambda\right)\sin\left(k\Lambda\right)}{k\Lambda} \right\}+ \\
& \frac{g\,\pi\,T^2}{6\,\kF k}\left\{ \Lambda\mu^2 \left[
  \frac{\cos\left[\left(\kF-k\right)\Lambda\right]}{\kF-k}-\frac{\cos\left[\left(\kF+k\right)\Lambda\right]}{\kF+k}\right] \right.+ \\
& \hspace{1.2cm}\left. \frac{\left(\kF^2-\kF k-\mu^2\right)\sin \left[\left(\kF-k\right)\Lambda\right]}{\left(\kF-k\right)^2} - \frac{\left(\kF^2+\kF k-\mu^2\right)\sin \left[\left(\kF+k\right)\Lambda\right]}{\left(\kF+k\right)^2}\right\}\;.
\end{split}
\end{equation}
Analytic expressions for the critical temperatures are given by 
\begin{equation}\label{eq:Tcrit0}
 T^2_{c,0}(\nu)=\frac{6\,\kF^4}{\pi^2}\frac{\left[\sin(\B{\nu})-\Si{\B{\nu}}\right]}{\left[ \left(2\mu^2-\kF^2\right)\B{\nu}\cos(\B{\nu}) + \left(\mu^2\B{\nu}^2+\kF^2-2\mu^2\right)\sin(\B{\nu})\right]}\;,
\end{equation}
\begin{equation}\label{eq:TcritFermi}
\begin{split}
 T^2_{c,\mathrm{F}}(\nu)=& \frac{24\, \kF^4}{\pi^2} \frac{\left[1-\cos(2\B{\nu})-\B{\nu}\Si{2\B{\nu}}\right]}{
 \B{\nu}^2\left[4\kF^2-2\mu^2\cos(\B{\nu})\right]+\B{\nu}\left[  \left(\mu^2-2\kF^2\right)\sin(2\B{\nu})\right]}\;,
\end{split}
\end{equation}
\begin{equation}\label{eq:TcritDFermi}
\begin{split}
 & T^2_{c,\mathrm{F}'}(\nu)= \frac{36\, \kF^4}{\pi^2} \left[2\B{\nu}^2 - 4 \sin(\B{\nu})^2 +\B{\nu} \sin(2\B{\nu})\right] \times \\
 & \left[ 4\mu^2 \B{\nu}^4 - 6 \B{\nu}^2 (2\kF^2+(\kF^2-2\mu^2)\cos(2\B{\nu})) + 3 (3\kF^2- 2\mu^2) \B{\nu} \sin(2\B{\nu}) + 6 \mu^2 \B{\nu}^3 \sin(2\B{\nu})\right]^{-1}\;.
\end{split}
 \end{equation}
Here $\B{\nu}$ are the zeros of the Bessel function $J_{3/2}$ typically called $j_{3/2,\nu}$ and $\mathrm{Si}$ is the integral sine.
Note that $T^2_{c,0}(\nu)>0$ only for odd $\nu$, while $T^2_{c,\mathrm{F}}(\nu),T^2_{c,\mathrm{F}'}(\nu)<0$ for all $\nu$.

\twocolumngrid
\bibliographystyle{utphys}
\bibliography{Orbifold}

\end{document}